\documentclass[conference,10pt]{IEEEtran}
\usepackage[ruled,vlined]{algorithm2e}
\usepackage{amsmath}
\usepackage{amssymb}
\usepackage{cite}
\usepackage{color}
\usepackage{float}
\usepackage[T1]{fontenc}
\usepackage{graphics}
\usepackage{epsfig}
\usepackage{microtype}
\usepackage[caption=false]{subfig}
\usepackage{textcomp}
\restylefloat{table}
\usepackage{setspace}

\begin{document}
\title{Combining Contention-Based Spectrum Access and Adaptive Modulation using Deep Reinforcement Learning}
\author{
\IEEEauthorblockN{Akash Doshi and  Jeffrey G. Andrews}
\IEEEauthorblockA{Department of Electrical and Computer Engineering, University of Texas at Austin, TX 78712}}
\maketitle 
\normalsize
\begin{abstract}
The use of unlicensed spectrum for cellular systems to mitigate spectrum scarcity has led to the development of intelligent adaptive approaches to spectrum access that improve upon traditional carrier sensing and listen-before-talk methods.  We study decentralized contention-based medium access for base stations (BSs) of a single Radio Access Technology (RAT) operating on unlicensed shared spectrum.  We devise a distributed deep reinforcement learning-based algorithm for both contention and adaptive modulation, modelled on a two state Markov decision process, that attempts to maximize a network-wide downlink throughput objective. Empirically, we find the (proportional fairness) reward accumulated by a policy gradient approach to be significantly higher than even a genie-aided adaptive energy detection threshold. Our approaches are further validated by improved sum and peak throughput. The scalability of our approach to large networks is demonstrated via an improved cumulative reward earned on both indoor and outdoor layouts with a large number of BSs.
\end{abstract}


\maketitle
\section{Introduction}
Unlicensed spectrum is set to play an increasing role in delivering broadband data traffic, both via WiFi and cellular standards-compliant systems.  For example, over one GHz of bandwidth including the entire 6 GHz band was released for unlicensed use by the FCC in 2020. An early but still recent example of cellular use of unlicensed spectrum is the deployment of Long Term Evolution (LTE) in the unlicensed 5 GHz band, known as LTE-U, licensed-assisted access (LAA), and other proprietary names \cite{patriciello2020nr}\cite{qualcommlteu2014}, which allowed peak rates of advanced LTE devices to approach or even exceed 1 Gbps.  Unlike licensed spectrum, access to unlicensed spectrum is free, but regulated by spectrum sensing technologies such as Listen-Before-Talk \cite{etsi}, which requires a transmitter to perform a Clear Channel Assessment before accessing spectrum. 

Recently, multi-agent reinforcement learning (RL) techniques have been utilized to improve unlicensed spectrum access \cite{lunden2011reinforcement,naparstek2018deep,tonnemacher2019machine}, however they do not model the asynchronous nature of contention-based access and have not been extended to perform adaptive modulation and coding (AMC). 
Meanwhile, several papers have utilized RL to design a rule for performing AMC \cite{leite2012flexible,bruno2014robust,mota2019adaptive}, however these are single-agent RL algorithms not concerned with cooperative spectrum sharing. In \cite{Dosh2101:Deep}, we developed a distributed Deep Q Network (DQN) based spectrum sharing algorithm incorporating contention-based medium access and adapted Proximal Policy Optimization (PPO) \cite{schulman2017proximal} to the aforementioned setting in \cite{dosh2021proximal}. In this paper, we provide a nontrivial generalization of \cite{Dosh2101:Deep} and \cite{dosh2021proximal}, by expanding the action space of the BS from a simple yes/no transmit decision to a choice of modulation scheme, hence providing a framework for \textit{joint rate adaptation and medium access}. Moreover, we propose modifications to the training algorithm of \cite{Dosh2101:Deep} that provide for scalability of the approach to more practical scenarios with a large number of BSs. 

\section{System Model \& Problem Statement} \label{sec:system_model}
\subsection{Mathematical Formulation} \label{subsec:math_form}
Consider a downlink cellular deployment of $N$ BSs, with a single UE scheduled per time slot per BS. Assume that the DL transmission for all BSs occurs on the same sub-band of shared unlicensed spectrum. With the same UE scheduled at a given BS for $L$ consecutive time slots, the medium access control (MAC) algorithm at the BS has to determine whether or not to transmit in each time slot, and if it chooses to transmit, which modulation scheme to use.

Assume each BS has a set of $K$ modulation schemes $M = \{M_1, M_2 \ldots M_K\}$ to choose from. Denote the transmit decision of BS $i$ by $a_i \in \{0,1\}$ and the transmit modulation scheme by $m_i \in M$. To recover the symbols $s_j$ transmitted by BS $j$, UE $j$ performs least squares (LS) equalization, assuming perfect knowledge of the channel between BS and UE $j$, followed by Maximum Likelihood decoding. Denote the symbol error probability of the recovered burst of symbols at UE $j$ by $P_{s,j}$. The throughput $R_j[n]$ achieved on a channel of bandwidth $W$ in time slot $n$ could then be approximated by
\begin{equation} \label{eq:Rate}
    R_{j}[n] = W(1 - P_{s,j}[n])a_j[n]\log_2(m_j[n]).
\end{equation}
We do not consider coding (which would introduce a multiplicative coding rate term $c_j[n]$ in \eqref{eq:Rate}), and only consider adaptive modulation. This is only for ease of implementation, and we compensate for this by considering both square and non-square (cross) QAM constellations to improve the granularity of $R_j[n]$, as will be highlighted in Section \ref{sec:Results}.  Denoting by $\overline{X}_j[n]$ the exponentially smoothed average rate seen at UE $j$, we have
\begin{equation}
    \overline{X}_j[n] = (1-1/\tau) \overline{X}_j[n-1] + (1/\tau)R_j[n]
\end{equation}
where $\tau > 1$ is the smoothing coefficient. Our objective is to maximize the long term average rate $\lim_{n\rightarrow \infty} \overline{X}_j[n]$ of each UE $j$ by choosing the appropriate transmit decision $a_j[n]$ and modulation scheme $m_j[n]$ at BS $j$ in each time slot $n$ in a decentralized fashion. As a benchmark, we will utilize the centralized proportional fairness (PF)-based BS scheduler that computes the optimal rate vector $\mathbf{R}^*[n]$ in each time slot as
\begin{equation} \label{eq:iterative_scheduler}
    \mathbf{R}^*[n] = \underset{\mathbf{R}[n]}{\mathrm{arg\ max\ }} \sum_{j=1}^{N} \frac{R_j[n]}{\overline{X}_j[n]}.
\end{equation}
In a given time slot $n$, determining $\mathbf{R}^*[n]$ is equivalent to determining $\mathbf{a}^* \odot \mathbf{m}^* = \{a^*_i[n],m^*_i[n]\} ~ \forall i \in [N]$, which involves a combinatorial search over $(|K|+1)^N$ possible vectors, making it computationally infeasible in practice.

\subsection{Contention-based Medium Access}
We consider a simplified contention-based access mechanism in which each time slot is divided into a contention and data transmission phase, with each contention phase further divided into $N$ mini-slots. At the beginning of each time slot, each BS $i$ is allotted a random counter $\theta_i \in \{0, \ldots, N-1\}$, with $\theta_i$ decreasing by 1 every mini-slot. Hence, each BS can be in either of two states - waiting for its counter $\theta_i$ to expire in the End-Of-Slot (EOS) state or transmitting data after performing contention in the Contention (CON) state \cite{Dosh2101:Deep}. Denoting by $\mathbf{G}[n] $ the downlink $N \times N$\ interference channel matrix in time slot $n$, it follows from \eqref{eq:iterative_scheduler} that the state of the system is given by $\mathbf{s}[n] \triangleq \langle \mathbf{\overline{X}}[n-1],\mathbf{G}[n]\rangle$. However, each BS can only observe a relatively small part of $\mathbf{s}[n]$, depending on whether it is in the EOS or CON state. We assume that UE $i$ feeds back its received signal power $S_i$ and measured interference power $I_i$ to BS $i$ in the uplink (error-free), and that BS $i$ keeps track of the average rate of the UE it serves, $\overline{X}_i$. Hence we define the observations $\mathbf{o}^{\mathrm{EOS}}_i$ and $\mathbf{o}^{\mathrm{CON}}_i$ as 
\begin{align}
    \mathbf{o}^{\mathrm{EOS}}_i[n] &= \langle \overline{X}_i[n-1], S_i[n-1], I_i[n-1] \rangle \\ \mathbf{o}^{\mathrm{CON}}_i[n] &= \langle \mathbf{o}^{\mathrm{EOS}}_i[n], \mathcal{E}_i^{\theta_i}[n], \theta_i[n] \rangle,
\end{align}
where $\mathcal{E}_i^{\theta_i} = \{\mathcal{E}_{ij}^{\theta_{i}}\}_{j\in[N]}$ is the inter-BS energy vector at BS $i$, with $\mathcal{E}_{ij}^{\theta_{i}}$ being the energy received at BS $i$ from an ongoing transmission between BS $j$ and UE $j$. While traditional AMC does not utilize $\mathcal{E}_i^{\theta_i}[n]$, traditional spectrum sensing does not utilize $\mathbf{o}^{\mathrm{EOS}}_i[n]$. Hence a combination of the two in $\mathbf{o}^{\mathrm{CON}}_i[n]$ is aimed at improving both medium access and rate adaptation.

In order to acheive long term PF, for every transition from CON to EOS, each BS receives a common reward $r^{\mathrm{CON}} = \sum_{j=1}^{N} r_j[n] \ \forall \ n > 0$ and $r^{\mathrm{CON}}[0] = \sum_{j=1}^{N} \log(\overline{X}_j[0])$ where 
\begin{equation} \label{eq:per_ts_rwd}
    r_j[n] = \log\Bigg((1-1/\tau)\Big(1 + \frac{R_j[n]}{(\tau-1)\overline{X}_j[n-1]}\Big)\Bigg).
\end{equation}
Maximizing $\sum_{n=0}^L \gamma^n r[n]$ for large $L$ was shown to be equivalent to maximizing $\sum_{j=1}^{N}\log(\overline{X}_j[n])$ for $n \rightarrow \infty$ and $\gamma \rightarrow 1$ \cite{Dosh2101:Deep}, which in turn ensures long term PF \cite{wang2010scheduling}.

\section{Deep RL for Adaptive Medium Access \& Modulation}
Given a state $s$ and an action $a$, we have three terms associated with a typical single agent RL problem: $\pi(a|s;\Theta)$, $\mathcal{Q}^{\pi}(s,a)$ and $V^{\pi}(s)$. We denote by $\pi(a|s;\Theta)$ a policy parameterized by $\Theta$ that returns the probability of an agent selecting action $a$ in state $s$. If an agent starts from state $s$, chooses action $a$ and thereafter follows $\pi$, the expected reward accumulated is represented by $\mathcal{Q}^{\pi}(s,a)$. Finally, $V^{\pi}(s)$ denotes the expected reward accumulated by an agent following $\pi$ starting from state $s$.

\subsection{Adapting Independent DQN}
Define two $\mathcal{Q}$ networks at each BS $i$, $\mathcal{Q}_i^{\mathrm{EOS}}$ and $\mathcal{Q}_i^{\mathrm{CON}}$. Utilizing the Bellman optimality equation, we can write the sampled $\mathcal{Q}$-value updates for $\mathbf{o}^{\mathrm{EOS}}_i$ and $\mathbf{o}^{\mathrm{CON}}_i$ as
\begin{align}
     &\mathcal{Q}_i^{\mathrm{EOS}}(\mathbf{o}_i^{\mathrm{EOS}}) = \gamma^{\frac{1}{2}} \max_{a_i^{\mathrm{CON}}} \mathcal{Q}_i^{\mathrm{CON}}(\mathbf{o}_i^{\mathrm{CON}},a_i^{\mathrm{CON}}) \label{eq:q_eos}\\
    &\mathcal{Q}_i^{\mathrm{CON}}(\mathbf{o}_i^{\mathrm{CON}},a_i^{\mathrm{CON}}) = r^{\mathrm{CON}} + \gamma^{\frac{1}{2}} \mathcal{Q}_i^{\mathrm{EOS}}(\mathbf{o}_i^{\mathrm{EOS}}), \label{eq:q_con}
\end{align}
where $r^{\mathrm{EOS}}_i = a^{\mathrm{EOS}}_i = 0$ and $a_i^{\mathrm{CON}}=a_im_i$. Since $\mathcal{Q}_i^{\mathrm{EOS}}$ and $\mathcal{Q}_i^{\mathrm{CON}}$ only have access to local observations $\mathbf{o}^{\mathrm{EOS}}_i$ and $\mathbf{o}^{\mathrm{CON}}_i$ and also are not aware of $\mathbf{a}\odot\mathbf{m} - \{a^{\mathrm{CON}}_i\}$, we compensate for this partial observability by utilizing \textit{recurrent} DQNs \cite{hausknecht2015deep} in order to map observation histories $\langle \vec{o}_i, \vec{\mathcal{E}}_i^{\theta_i}, \vec{\theta}_i \rangle$ to actions. The DQN based algorithm is presented in Algorithm \ref{algo:dqn}.
\begin{algorithm} \label{algo:dqn}
\setstretch{1}
\SetAlgoNoLine
\SetAlgoHangIndent{0pt}
\DontPrintSemicolon
\For{$\mathrm{iteration}=1,2, \ldots $}{
    \For{$\mathrm{actor}=1,2, \ldots, N_{\mathrm{batch}}$}{
        Generate an episode of $L$ time slots.\;
        In each time slot, each BS $i$ chooses a transmit modulation scheme from $M'=\{0,M\}$ randomly w.p. $\epsilon$ or greedily based on $\mathcal{Q}_i^{\mathrm{CON}}$ w.p. $1 - \epsilon$.\;
    }
    \For{$i = 1,2, \ldots, N$}{
    Compute labels $\mathcal{L}_i^{\mathrm{EOS}}(\mathbf{s}^{\mathrm{EOS}})$  and $\mathcal{L}_i^{\mathrm{CON}}(\mathbf{o}_i^{\mathrm{CON}},a_i^{\mathrm{CON}})$ using \eqref{eq:q_eos} and \eqref{eq:q_con} at each time for all actors in parallel.\;
    Perform 1 epoch of gradient descent with batch size $N_{\mathrm{batch}}\times L$ on MSE loss to update weights of $\mathcal{Q}_i^{\mathrm{EOS}}$ and $\mathcal{Q}_i^{\mathrm{CON}}$.\; }
}
\caption[caption]{Spectrum Sharing Deep Q Learning}
\end{algorithm}

\subsection{Adapting Proximal Policy Optimization}
Define $N$ loss functions $L^{\mathrm{PPO}}_{\mathrm{EOS-CON}}(\Theta_i^{\pi_{\mathrm{CON}}},\vartheta_i^{V_{\mathrm{CON}}},\vartheta_i^{V_{\mathrm{EOS}}})$ corresponding to each BS $i$ \cite{dosh2021proximal}. In order to compute the loss functions, we modify the PPO update equations from \cite{schulman2017proximal} to incorporate the contention-based 2-state MDP as follows
\begin{multline}
     V_i^{\mathrm{target,EOS}} = V_i^{\mathrm{EOS}}(\mathbf{o}_i^{\mathrm{EOS}}) + \delta_{i,n}^{\mathrm{EOS}} + (\gamma^{\frac{1}{2}}\lambda)\delta_{i,n}^{\mathrm{CON}} \\+ \ldots + (\gamma^{\frac{1}{2}}\lambda)^{L-n+1}\delta_{i,L-1}^{\mathrm{EOS}}  \label{eq:V_EOS}
\end{multline}
\begin{multline}
     V_i^{\mathrm{target,CON}} = V_i^{\mathrm{CON}}(\mathbf{o}_i^{\mathrm{CON}}) + \delta_{i,n}^{\mathrm{CON}} + (\gamma^{\frac{1}{2}}\lambda)\delta_{i,n+1}^{\mathrm{EOS}} \\+ \ldots + (\gamma^{\frac{1}{2}}\lambda)^{L-n+1}\delta_{i,L-1}^{\mathrm{CON}} \label{eq:V_CON}
\end{multline}
\begin{equation}
    \hat{A}_i^{\mathrm{CON}} = V_i^{\mathrm{target,CON}}(\mathbf{o}_i^{\mathrm{CON}}) - V_i^{\mathrm{CON}}(\mathbf{o}_i^{\mathrm{CON}}), \label{eq:A_CON}
\end{equation}
where 
\begin{align} 
    \delta_{i,n}^{\mathrm{CON}} &= r[n] + \gamma^{\frac{1}{2}}V_i^{\mathrm{EOS}}(\mathbf{o}_i^{\mathrm{EOS}}[n+1]) - V_i^{\mathrm{CON}}(\mathbf{o}_i^{\mathrm{CON}}[n]) \label{eq:delta_CON}\\
    \delta_{i,n}^{\mathrm{EOS}} &= \gamma^{\frac{1}{2}}V_i^{\mathrm{CON}}(\mathbf{o}_i^{\mathrm{CON}}[n]) - V_i^{\mathrm{EOS}}(\mathbf{o}_i^{\mathrm{EOS}}[n]). \label{eq:delta_EOS}
\end{align}
The PPO based algorithm is presented in Algorithm \ref{algo:ppo}. Note that only $\pi_i^{\mathrm{CON}}$ is required for generating an episode, while the remaining NNs are employed only during training.
\begin{algorithm} \label{algo:ppo}
\setstretch{1}
\SetAlgoNoLine
\SetAlgoHangIndent{0pt}
\DontPrintSemicolon
\For{$\mathrm{iteration}=1,2, \ldots $}{
    \For{$\mathrm{actor}=1,2, \ldots, N_{\mathrm{batch}}$}{
        Generate an episode of $L$ time slots. \;
        In each time slot, each BS $i$ chooses not to transmit $/$ transmit with modulation scheme $M_k$ with probability $\pi_i^{\mathrm{CON}}[0] ~/~ \pi_i^{\mathrm{CON}}[k]$. \;
    }
    \For{$i = 1,2, \ldots, N$}{
    Compute target values $V_i^{\mathrm{target,EOS}}$,$V_i^{\mathrm{target,CON}}$ and advantage estimate $\hat{A}_i^{\mathrm{CON}}$ using  \eqref{eq:V_EOS}, \eqref{eq:V_CON} and \eqref{eq:A_CON} respectively at each time for all actors in parallel. \;
    By performing gradient ascent on $L^{\mathrm{PPO}}_{\mathrm{EOS-CON}}(\Theta_i^{\pi_{\mathrm{CON}}},\vartheta_i^{V_{\mathrm{CON}}},\vartheta_i^{V_{\mathrm{EOS}}})$, update weights of $\pi_i^{\mathrm{CON}}$, $V_i^{\mathrm{CON}}$ and $V_i^{\mathrm{EOS}}$. Perform 1 epoch with batch size $N_{\mathrm{batch}}\times L$}
}
\caption[caption]{Spectrum Sharing PPO}
\end{algorithm}

\subsection{Decentralized Actor Centralized Critic} \label{subsec:dacc}
Naive policy gradient methods perform poorly even in simple multi-agent settings \cite{NIPS2017_68a97503}. For instance, in the current setting, via recurrent neural networks, BS $i$ selects an action $a_i^{\mathrm{CON}}$ based on the policy $\pi_i^{\mathrm{CON}}: \langle \vec{o}_i, \vec{\mathcal{E}}_i^{\theta_i}, \vec{\theta}_i \rangle \rightarrow \{0,M\}$. If the learned policy is conditioned only on $\langle \vec{o}_i, \vec{\mathcal{E}}_i^{\theta_i}, \vec{\theta}_i \rangle$, it will exhibit high variability, increasing the variance of the gradients. Consequently, to stabilize the training and improve the learnt policy, we incorporate a \textit{decentralized actor centralized critic} approach, first proposed in \cite{NIPS2017_68a97503}. To this effect, we change the input to $V_i^{\mathrm{EOS}}$, $\mathcal{Q}_i^{\mathrm{EOS}}$ and $V_i^{\mathrm{CON}}$ by replacing $\mathbf{o}_i^{\mathrm{EOS}}$ with $\mathbf{s}^{\mathrm{EOS}}$ at each BS $i$. While we defined $\mathbf{o}_i^{\mathrm{EOS}}[n+1] = \langle \overline{X}_i[n], S_i[n], I_i[n] \rangle$, we have $\mathbf{s}^{\mathrm{EOS}}[n+1] = \langle \overline{\mathbf{X}}[n], \mathbf{S}[n], \mathbf{I}[n] \rangle$ i.e. it will contain the average rate, signal and interference power of all UEs in the previous time slot. Hence the input to $V_i^{\mathrm{CON}}$ will be $\langle \mathbf{s}^{\mathrm{EOS}}, \mathcal{E}_i^{\theta_i},\theta_i \rangle$, while the input to $\pi_i^{\mathrm{CON}}$ and $\mathcal{Q}_i^{\mathrm{CON}}$ remains unchanged. This training modification enabled scalability to practical environments with large number of BSs.

\section{Simulation Details}
\subsection{Channel Modelling and Modulation Schemes}
\begin{figure}
\centering
    \subfloat[Indoor office layout]{\includegraphics[width = 2.8in]{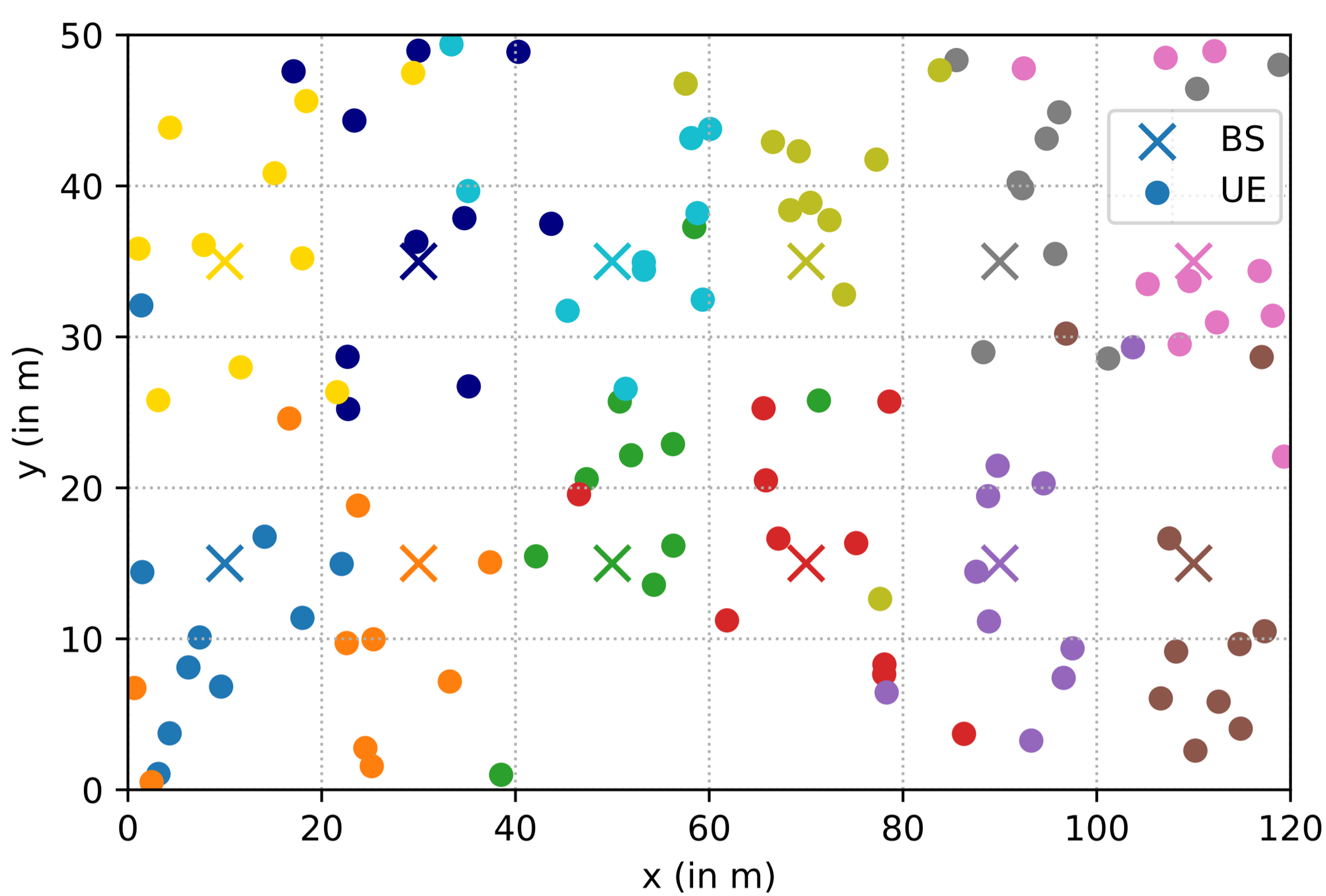}
    \label{fig:office_layout}}
    \hspace{0.02in}
    \subfloat[UMi-Street Canyon layout]{\includegraphics[width=2.4in]{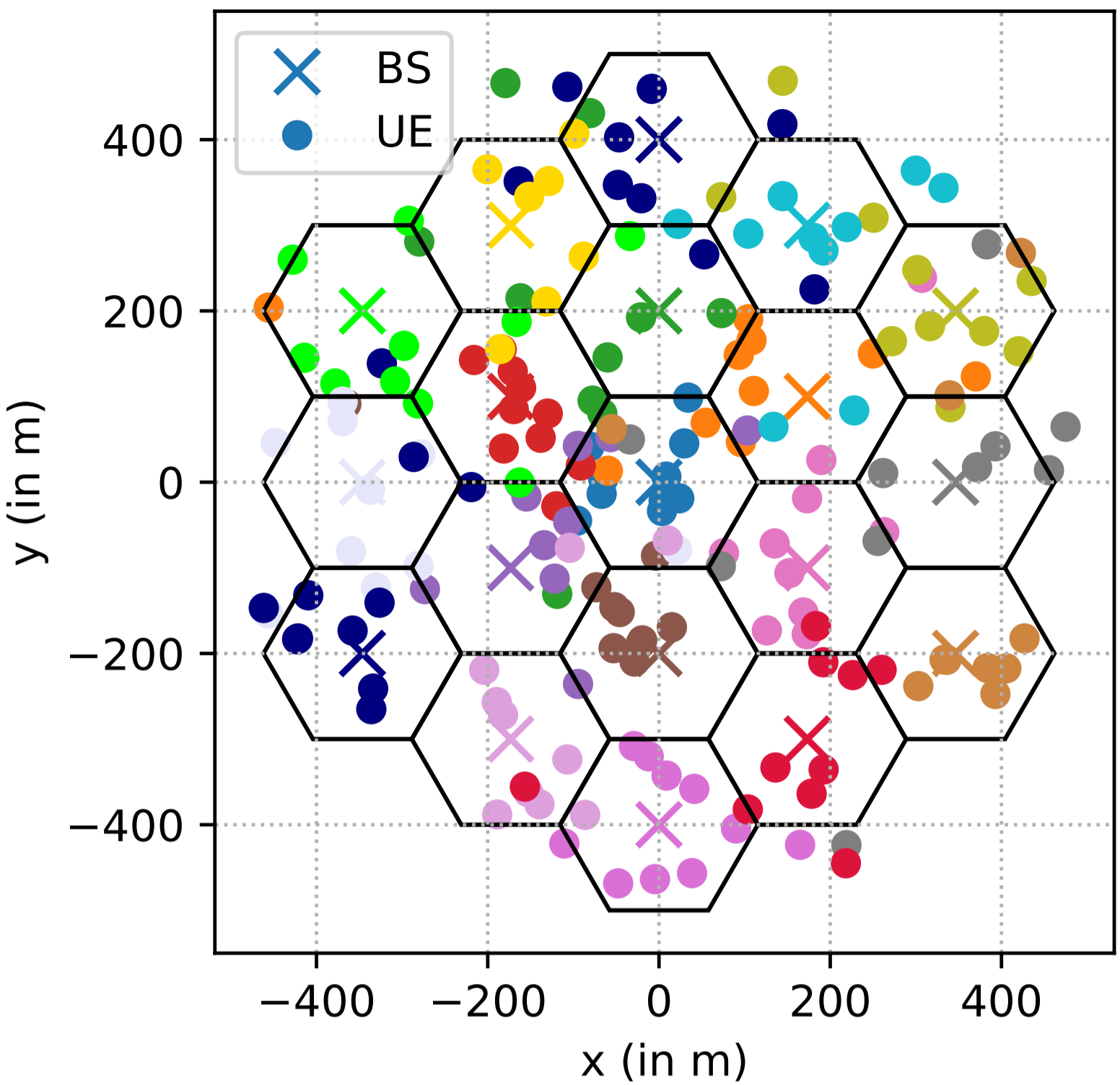}
    \label{fig:layout_hexagon}}
\caption{3GPP InH-Office and UMi-Street Canyon layout used for simulations.}
\label{fig:standards_layout}
\end{figure}

We consider two deployment scenarios from \cite{3gpp.38.901} - an indoor and an outdoor layout. The indoor hotspot deployment (InH-Office), as depicted in Fig. \ref{fig:office_layout}, has 12 BS's mounted at a height of 3m on the ceiling. The outdoor layout is an Urban Micro (UMi) layout, as depicted in Fig. \ref{fig:layout_hexagon}, where 19 BSs are distributed in a hexagonal layout at a height of 10m. The pathloss (PL) model between nodes (BS and UEs) captures Line-Of-Sight(LOS)/ Non-Line-Of-Sight(NLOS) properties of a link, frequency dependent path loss for LOS/ NLOS links and shadowing as part of large-scale fading parameters. The center frequency used for modelling is $6$ GHz. The temporal evolution of the channel coefficients is modelled as a slow fading channel using a first order IIR filter with fading coefficient $\alpha=0.1$.

We set $M = \{4, 8, 16, 32, 64, 128, 256\}$. For 
(i) M-PSK constellation $M_k = \{8\}$, its symbol $s$ is given by 
\begin{equation}
    \mathrm{exp}(2\pi j \beta/M_k), ~~~~ \beta = 0, \ldots, M_k - 1
\end{equation}
(ii) for square constellations $M_k \in \{4, 16, 64, 256\}$, the in-phase and quadrature components of its symbol $s$ independently takes values in the set 
\begin{equation}
    \bigg \{ \sqrt{\frac{3}{2(M_k-1)}} \big(2\beta + 1 - \sqrt{M_k} \big) \bigg \}, ~~~~ \beta = 0, \ldots, \sqrt{M_k} - 1
\end{equation}
(iii) and for cross QAM constellations $M_k \in \{32,128\}$, we define a block parameter $v = \sqrt{\frac{M}{32}}$ \cite{smith1975odd}. Cross-QAM constellation can then be constructed by a square block array with the 4 corner blocks deleted, each block with $v^2$ uniform distributed points \cite{zhang2010exact}. For all constellations, we ensure that the constellation points are normalized to unit power in expectation.

\subsection{PF Scheduler and ED Threshold Baselines}
Three baselines are used for benchmarking the performance of the RL PPO and DQN algorithms - centralized PF, genie-based adaptive ED and a single ED threshold of $E_0 = - 72 ~\mathrm{dBm}$. The centralized PF-based BS scheduler was presented in Section \ref{subsec:math_form}, and we utilize it to determine whether a BS will transmit in each time slot, by setting $R_j[n] = W\log_2({1 + \mathrm{SINR}_j[n]})$. The single ED threshold baseline allows a BS to transmit if $\sum_{j=1}^{N} \mathcal{E}_{ij}^{\theta_{i}} < E_0$. ``Adaptive ED'' finds the ED threshold that maximizes $\sum_{n=0}^{L}\gamma^n r[n]$ for the given configuration of UEs from a set of ED thresholds ranging from -22 dBm to -92 dBm. Using these baselines, each BS $i$ determines $a_i$. Suppose $a_i = 1$. Then to determine which modulation scheme $m_i$ is chosen by BS $i$, we solve  
\begin{equation} \label{eq:mod_scheme}
    m_i = \arg \max_j ||(1 - P_{s,j}) \log_2 M_j||^2_2
\end{equation}
where $P_{s,j}$ is approximated by \cite{goldsmith2005wireless} as
\begin{align} \label{eq:P_s}
    P_{s,j} = \begin{cases}1 - \Big(1 - \frac{2(\sqrt{M_j} -1)}{\sqrt{M_j}} Q\Big( \sqrt{\frac{3\cdot \mathrm{SINR}_i}{M_j - 1}}\Big) \Big)^2  \\
    2Q\Bigg(\sqrt{2\cdot \mathrm{SINR}_i \sin^2\Big({\frac{\pi}{M_j}}\Big)} \Bigg) \\
    4Q\Big( \sqrt{\frac{3\cdot \mathrm{SINR}_i}{M_j - 1}}\Big),
    \end{cases}
\end{align}
where the first branch is utilized for $M_j=\{4,16,64,256\}$, the second for $M_j = 8$ and the third for $M_j = \{32,128\}$. There are two key observations that can be made here - first, by utilizing \eqref{eq:mod_scheme} to compute $m_i$, we are assuming a genie in all baselines that can provide BS $i$ with the value of $\mathrm{SINR}_i$, once each BS has decided whether or not to transmit. No such genie is required by the RL algorithms. Second, if we were simply determining whether or not to transmit, then centralized PF would be an upper bound on the attainable reward \cite{wang2010scheduling}. However, with $K+1$ actions to choose from per BS, finding the modulation schemes that would maximize the PF metric would require $(K+1)^N$ simulations that compute $P_{s,j} ~\forall j$ per time slot which is not feasible. Hence, we continue to choose a binary decision $a_i$ from $2^N$ options for the PF scheduler at BS $i$, and then compute the modulation scheme $m_i$ using \eqref{eq:mod_scheme}.

\section{Results} \label{sec:Results}
We will refer to the position of the BSs as a layout, and the position of the UEs as a configuration. For both of the layouts, we randomly sample $2\times10^4$ configurations. From these, we randomly sample $N_{\mathrm{batch}} = 8$ configurations and generate one episode using each of the $N_{\mathrm{batch}}$ configurations at each training iteration. To plot the validation results, we evaluate the trained models obtained after every 10 iterations on 10 randomly sampled configurations and averaged over 10 realizations of each configuration. For every realization of each UE configuration, we also compute the sum rate $W \sum_{j=1}^N \overline{X}_j[L]$ and max rate $W \max_{j} \overline{X}_j[L]$ obtained using the RL PPO and DQN algorithms at the end of $L$ time-steps.

\begin{figure*}
    \centering
    \subfloat[Cumulative Reward]{\includegraphics[width=2.33in]{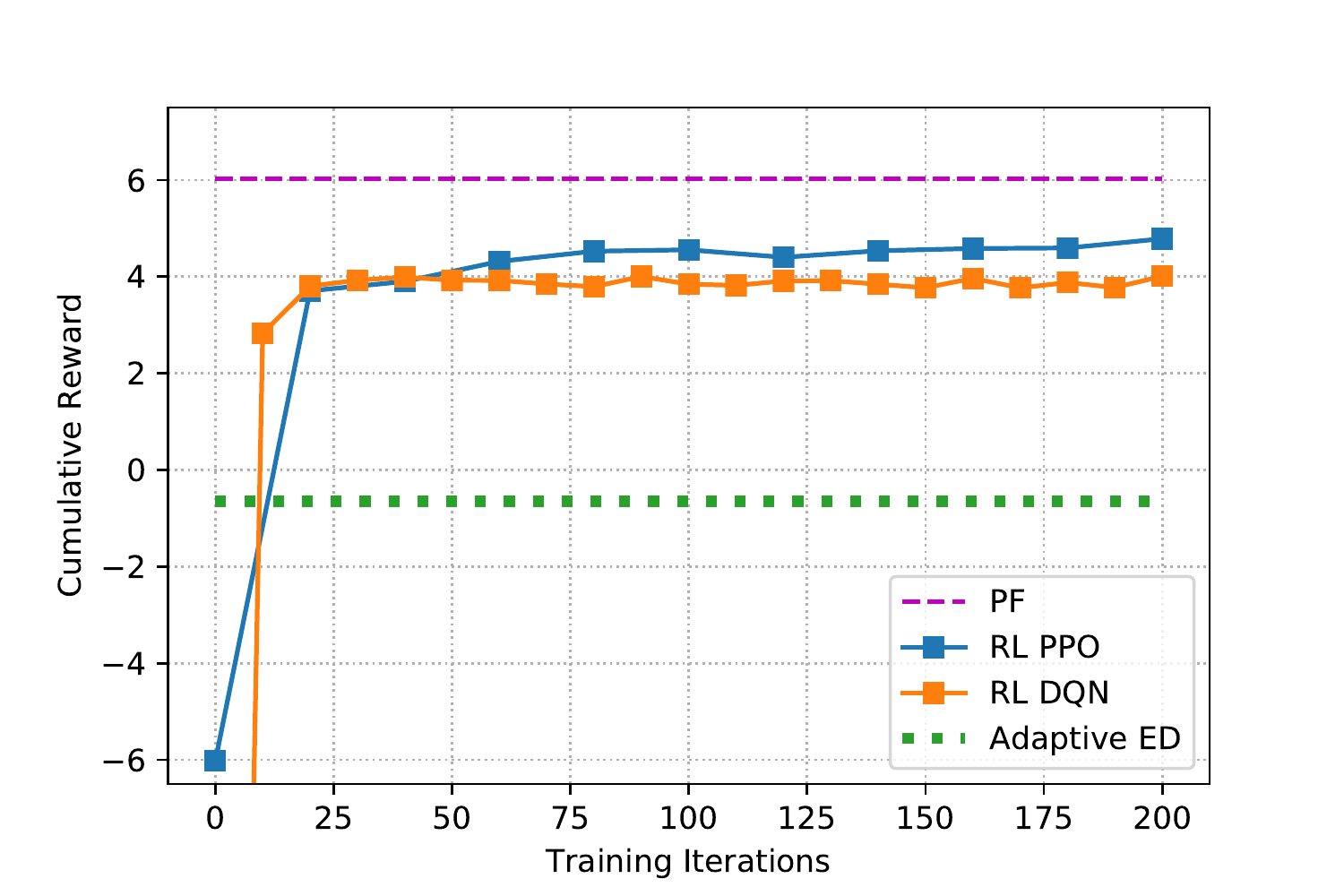}
    \label{fig:reward_iter_all}}
    \subfloat[UE throughput]{\includegraphics[width=2.33in]{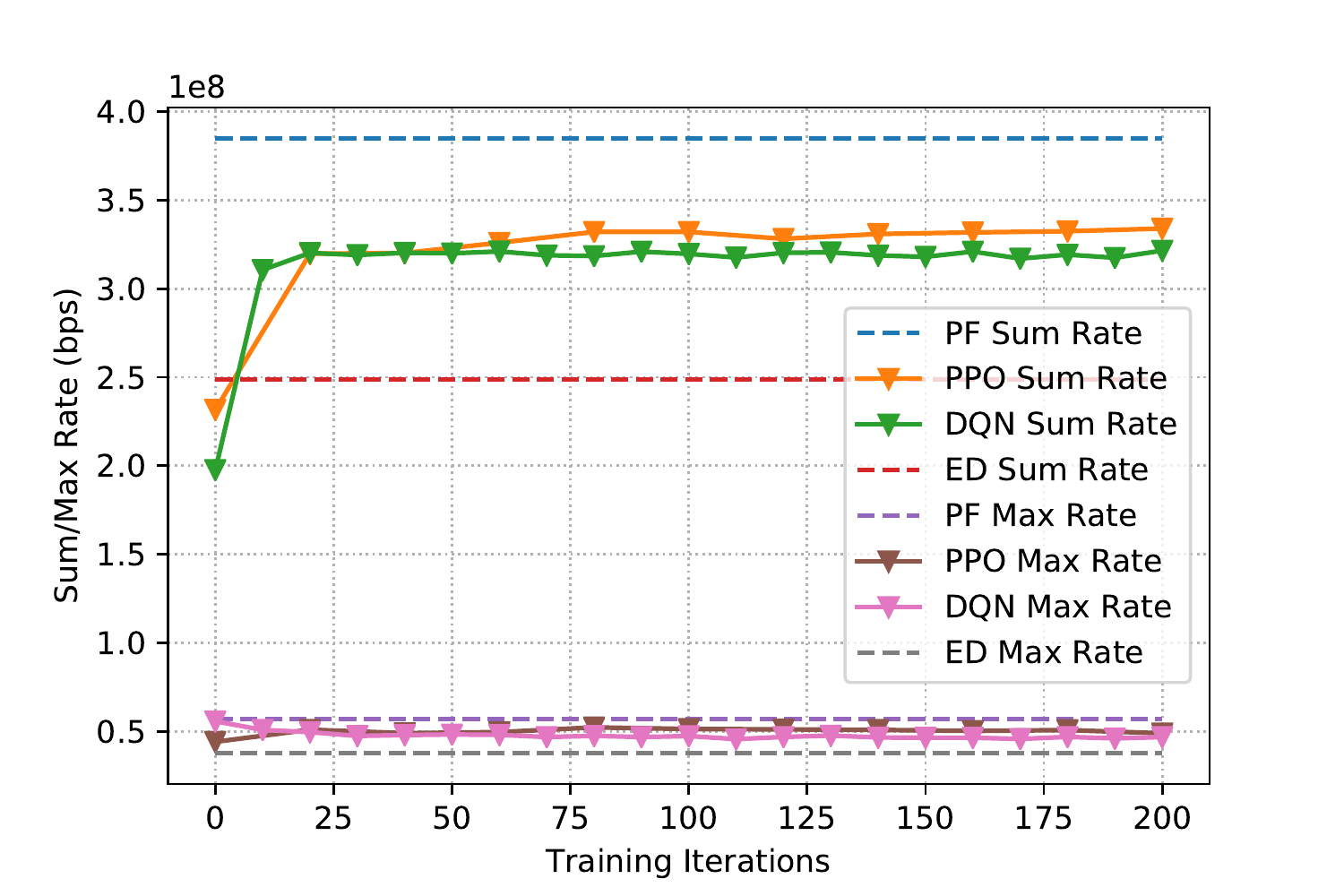}
    \label{fig:rate_iter_all}}
    \subfloat[Cumulative Reward with $\mathcal{E}_i^{\theta_i,3}$]{\includegraphics[width=2.33in]{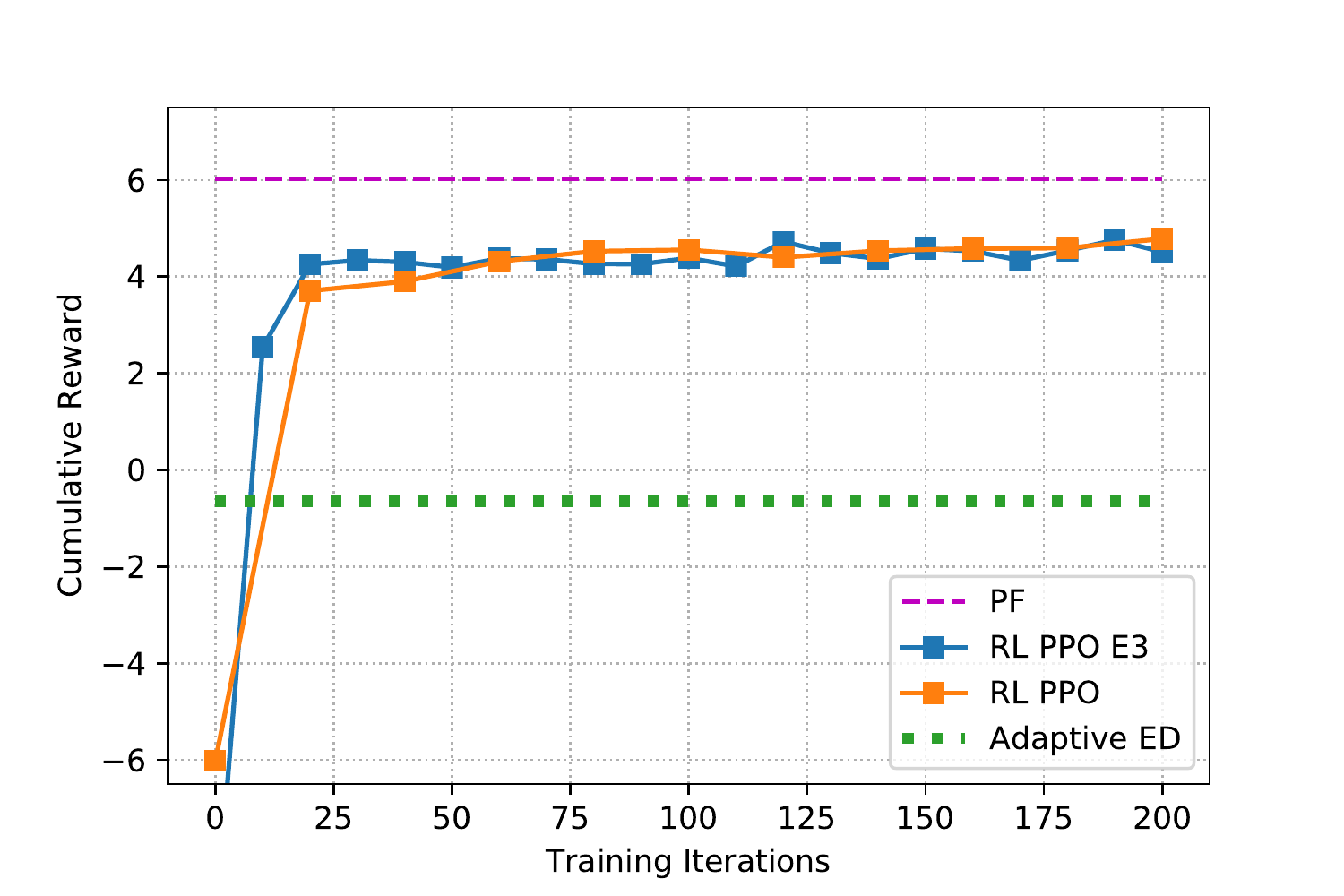}
    \label{fig:fixed_E_all}}
    \caption{Results evaluated on Validation Set for InH-Office layout.For \textit{Layout 1}, ED $= -72$ dBm has a reward of -22.03 and has not been shown.}
    \label{fig:results_Inh}
\end{figure*}

\begin{figure*}
    \centering
    \subfloat[Cumulative Reward]{\includegraphics[width=2.33in]{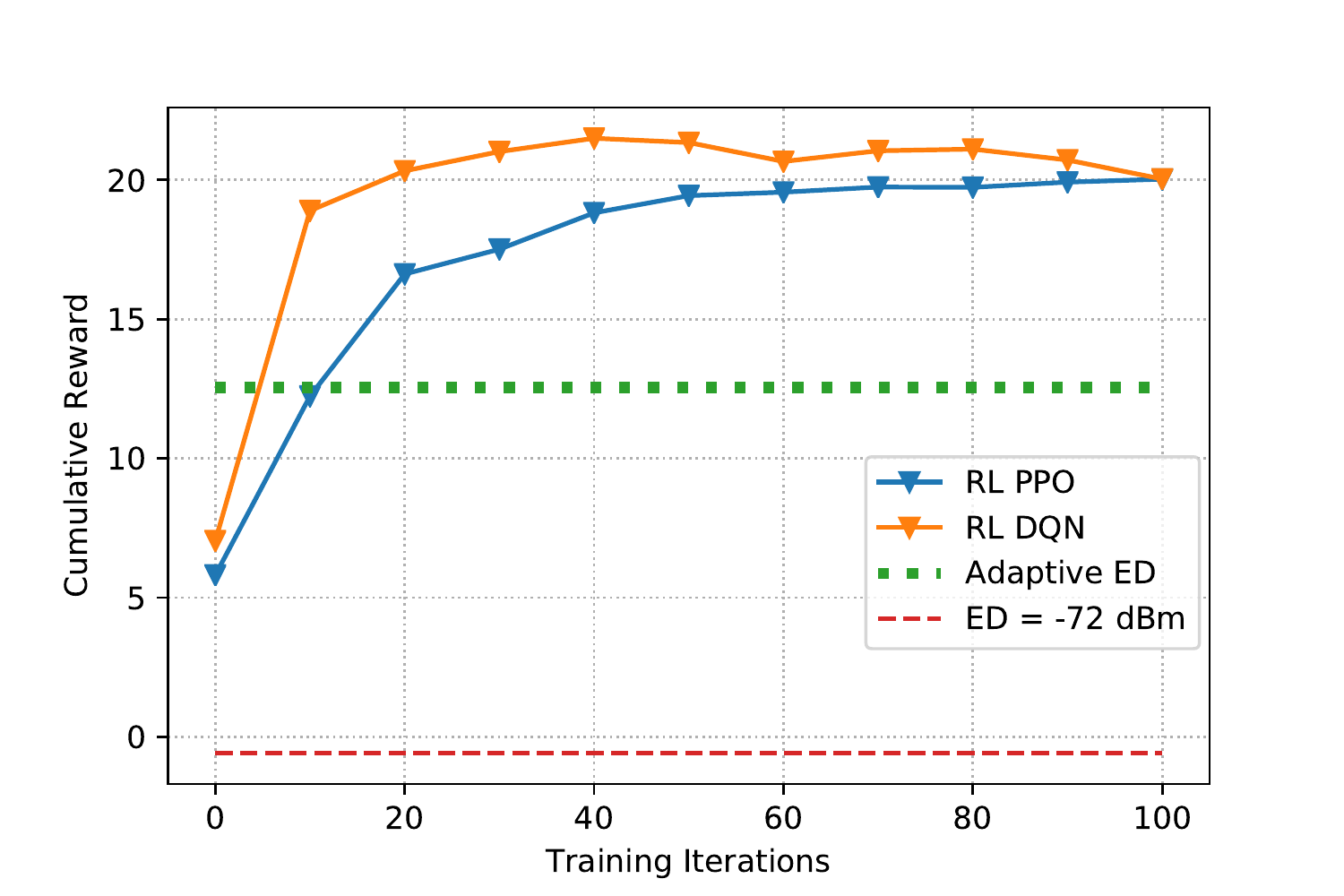}
    \label{fig:reward_iter_hexagon}}
    \subfloat[UE throughput]{\includegraphics[width=2.33in]{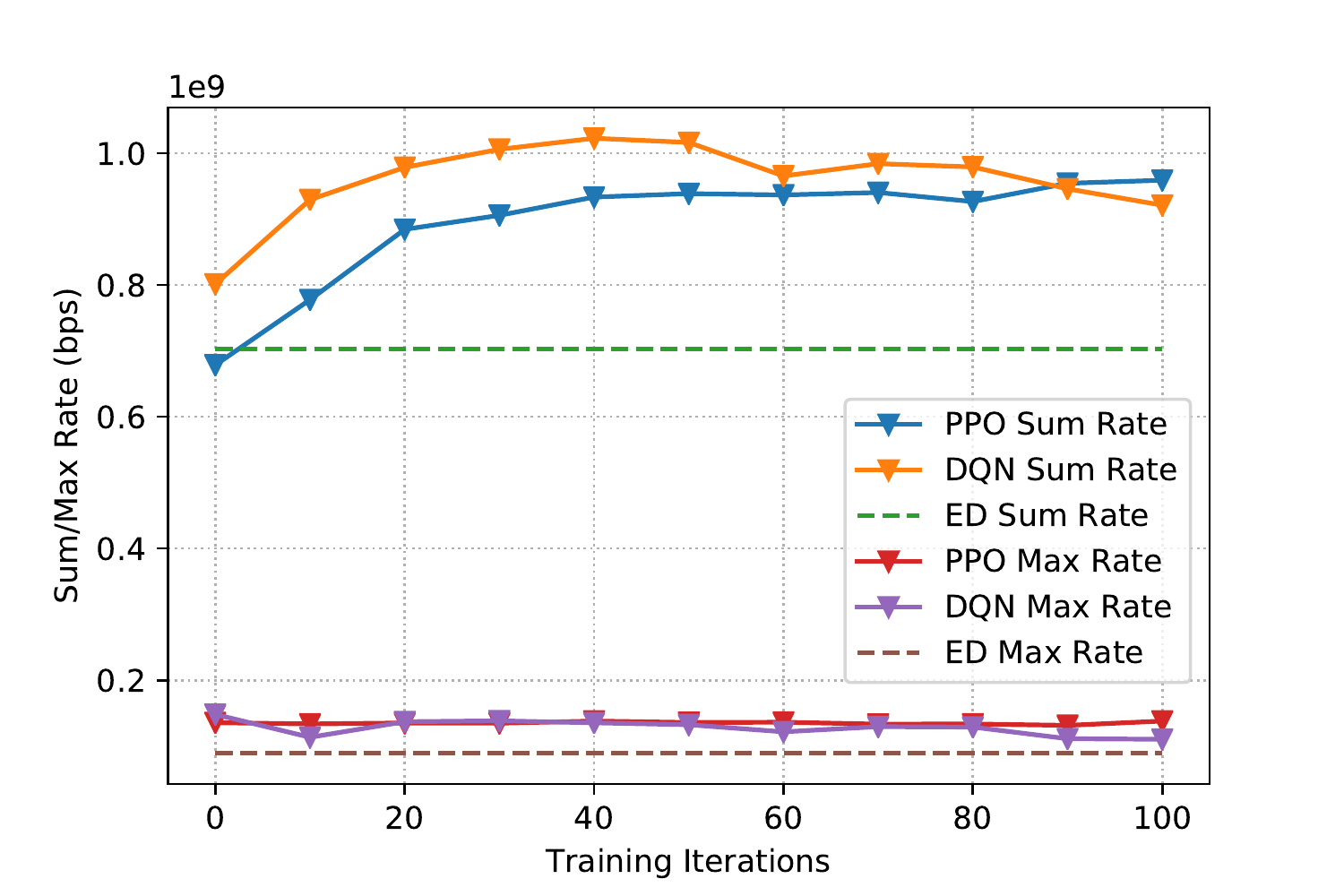}
    \label{fig:rate_iter_hexagon}}
    \subfloat[Cumulative Reward with $\mathcal{E}_i^{\theta_i,5}$]{\includegraphics[width=2.33in]{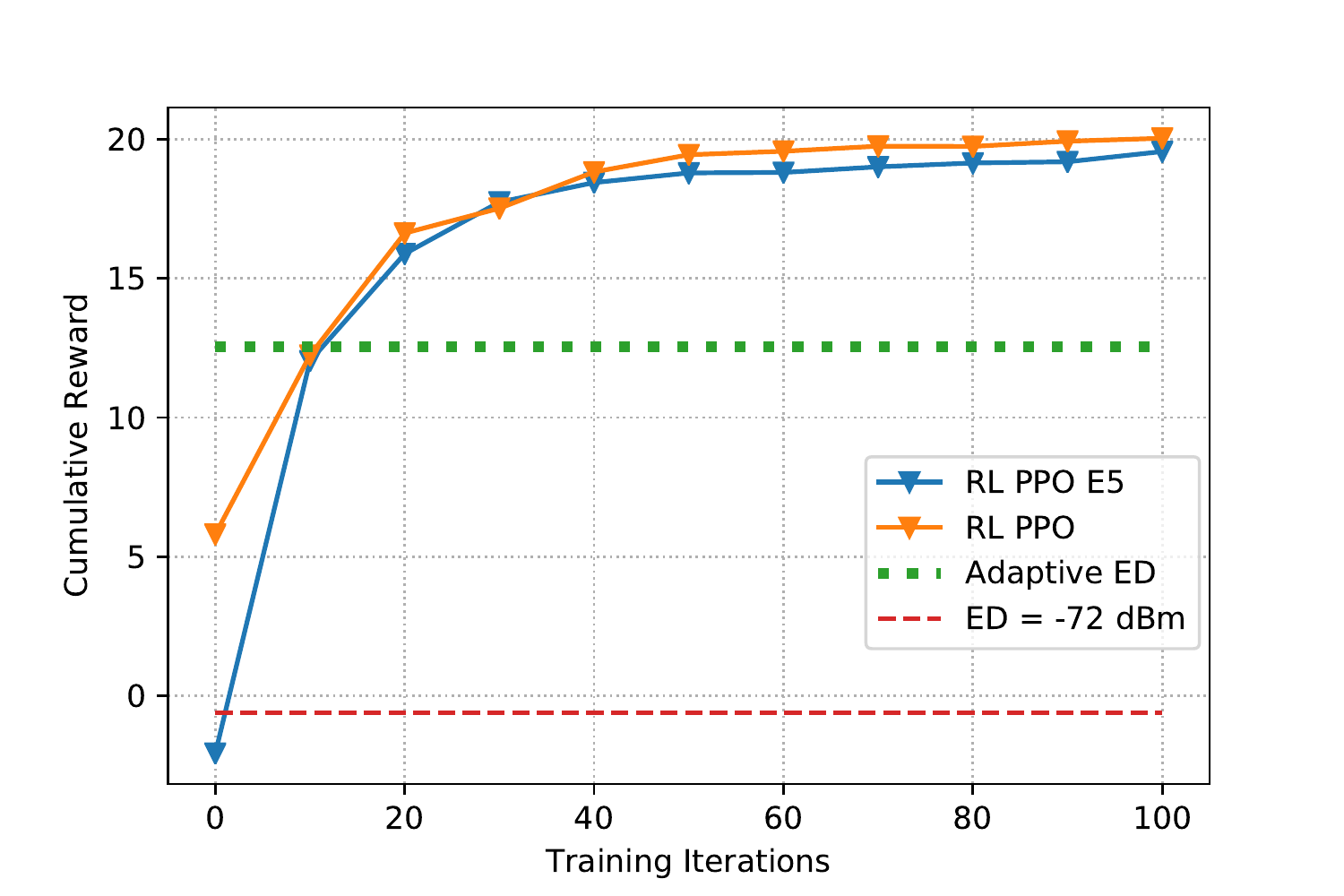}\label{fig:fixed_E_hexagon}}
    \caption{Results evaluated on Validation Set for UMi-Street Canyon layout.}
    \label{fig:results_Umi}
\end{figure*}

\textbf{RL significantly outperforms Adaptive ED:} In both layouts, the reward accumulated by the RL algorithms is significantly higher than even the genie-aided adaptive ED baseline as seen in Fig \ref{fig:reward_iter_all} and \ref{fig:reward_iter_hexagon}. This gain is also reflected in the significant increase in sum rate in Fig. \ref{fig:rate_iter_all} and \ref{fig:rate_iter_hexagon}. At the same time, we can see that ED $= -72 \mathrm{dBm}$ provides significantly varying degree of fairness in different scenarios, depending on the BS separation.

\textbf{Rapid Training Convergence and Low Sample Complexity:} Since we utilize a \textit{decentralized actor centralized critic} approach for training the NNs deployed at each BS, we were able to significantly lower the number of training iterations required for the cumulative reward to stabilize, while also lowering the number of training samples needed at each iteration as compared to \cite{Dosh2101:Deep}. For all layouts in this paper, we were able to acheive convergence within 100 iterations with a memory of size $N_{\mathrm{batch}} \times L = 1.6 \times 10^4$ samples at each iteration, while in \cite{Dosh2101:Deep}, each layout considered required atleast 4000 iterations to acheive training convergence with a memory size of $5 \times 10^5$ samples at each iteration.

\textbf{Truncating $\mathcal{E}_i^{\theta_i}$ does not impact RL performance:} One of the drawbacks of the inter-BS energy vector $\mathcal{E}_i^{\theta_i}$ introduced as part of the CON state in Section \ref{sec:system_model}, is that its length equals the number of BSs in the layout. This is clearly not scalable to any practical deployment. However, in \cite{Dosh2101:Deep}, we depicted the need to preserve the ordering of elements in $\mathcal{E}_i^{\theta_i}$ across agents, such that the CON NNs could assign an identity to the interfering transmitter. Hence we adopt the following strategy - restrict $\mathcal{E}_i^{\theta_i}$ to be of fixed length, not dependent on the number of BSs in the layout. Then append each energy measurement as $\langle j, \mathcal{E}_i^{\theta_{ij}}\rangle$ before inputting to the CON NNs $\pi_i^{\mathrm{CON}}$ / $V_i^{\mathrm{CON}}$. Adding the index $j$ is aimed at informing the NN as to the identity of the interfering BS. The validation results for the UMi Street Canyon layout are shown in Fig. \ref{fig:fixed_E_hexagon} with length of $\mathcal{E}_i^{\theta_i,5}$ as 5 for $N = 19$. It should be noted that the 5 entries in $\mathcal{E}_i^{\theta_i,5}$ are the 5 largest entries from $\mathcal{E}_i^{\theta_i}$. Similar results are shown for the InH-Office layout in Fig. \ref{fig:fixed_E_all} with length of $\mathcal{E}_i^{\theta_i,3}$ as 3 for $N = 12$. The results indicate that we can use a smaller number of entries in $\mathcal{E}_i^{\theta_i}$ without impacting the policy learnt, improving the practicality of the approach.

\section{Conclusions} \label{sec:conc}
In this paper, we formulated novel distributed implementations of two single agent RL algorithms - PPO \cite{schulman2017proximal} and DQN \cite{mnih2015human} - adapted to a contention-based medium access $\textit{DEC-POMDP}$ to choose the optimal transmit modulation scheme in each time slot that maximized the long term average throughput for all UEs. Our framework jointly utilized UE signalling parameters such as the signal and interference power from the last time slot in combination with spectrum sensing information from the current time slot to perform rate adaptation in a way that maximized proportional fairness network-wide. We utilized a centralized training procedure with decentralized inference -- \textit{decentralized actor centralized critic} -- to improve the scalability of our approach, applying our algorithms to both indoor and outdoor layouts of 12 and 19 BSs respectively. The RL algorithms were found to acheive a significantly higher PF metric than a genie-aided configuration adaptive ED threshold, and also improved the sum and max rates achieved by the UEs.

\bibliographystyle{IEEEtran}
\bibliography{bibtex.bib}

\end{document}